\documentclass[a4paper]{article}

\usepackage{INTERSPEECH_v2}

\title{Recognizing Multi-talker Speech with Permutation Invariant Training}
\name{Dong Yu$^1$, Xuankai Chang$^2$, Yanmin Qian$^2$ \thanks{Dong Yu and Yanmin Qian are the corresponding authors.} \thanks{1. Part of the work was done while at Microsoft Research.} \thanks{2. Xuankai Chang and Yanmin Qian was supported by the Shanghai Sailing Program No. 16YF1405300, the China NSFC projects (No. 61573241 and No. 61603252) and the Interdisciplinary Program (14JCZ03) of Shanghai Jiao Tong University in China. Experiments have been carried out on the PI supercomputer at Shanghai Jiao Tong University.}}
\address{
  $^1$Tencent AI Lab, Seattle, USA\\
  $^2$Department of Computer Science and Engineering, Shanghai Jiao Tong University, Shanghai, China}
\email{dongyu@ieee.org, xuank@sjtu.edu.cn, yanminqian@sjtu.edu.cn}

\begin{document}

\maketitle
\begin{abstract}
  In this paper, we propose a novel technique for direct recognition of multiple speech streams given the single channel of mixed speech, without first separating them. Our technique is based on permutation invariant training (PIT) for automatic speech recognition (ASR). In PIT-ASR, we compute the average cross entropy (CE) over all frames in the whole utterance for each possible output-target assignment, pick the one with the minimum CE, and optimize for that assignment. PIT-ASR forces all the frames of the same speaker to be aligned with the same output layer. This strategy elegantly solves the label permutation problem and speaker tracing problem in one shot.  Our experiments on artificially mixed AMI data showed that the proposed approach is very promising. 
\end{abstract}

\section{Introduction} \label{sec:intro}

Thanks to the significant progresses made in the recent years \cite{ASRBook-Yu2014,PretrainVSFineTune-Yu2010,CD-DNN-HMM-dahl2012,CD-DNN-HMM-SWB-seide2011,DNN4ASR-hinton2012,CNN4ASR-Abdel-Hamid2012,CNN-Trans-Abdel-Hamid2014,CLDNN-sainath2015,DeepCNN-bi2015,DeepCNN-qian2016.1,DeepCNN-qian2016.2,TFCNN-mitra2015,TDNN-peddinti2015,VGG-secru2016,Deepspeech2-amodei2015,FSMN-zhang2015,LACE-yu2016,HumanParity-Xiong2016,PIT-yu2017,PIT-Kolbak2017}, the ASR systems now surpassed the threshold for adoption in many real-world scenarios and enabled services such as Microsoft Cortana, Apple's Siri and Google Now, where close-talk microphones are commonly used.

However, the current ASR systems still perform poorly when far-field microphones are used. This is because many difficulties hidden by close-talk microphones now surface under distant recognition scenarios. For example, the signal to noise ratio (SNR) between the target speaker and the interfering noises is much lower than that when close-talk microphones are used. As a result, the interfering signals, such as background noise, reverberation, and speech from other talkers, become so distinct that they can no longer be ignored.

In this paper, we aims at solving the speech recognition problem when multi-talkers speak at the same time and only a single channel of mixed speech is available. Many attempts have been made to attack this problem. Before the deep learning era, the most famous and effective model is the factorial GMM-HMM \cite{FactorialHMM-ghahramani1997}, which outperformed human in the 2006 monaural speech separation and recognition challenge \cite{MonauralSpeechSepChallenge-Cooke2010}. The factorial GMM-HMM, however, requires the test speakers to be seen during training so that the interactions between them can be properly modeled. Recently, several deep learning based techniques have been proposed to solve this problem \cite{SingleChannelSep-Weng2015,DeepClustering-hershey2015,DeepClustering2-isik2016,AtrractorNet4SpeechSeparation-chen2017,PIT-yu2017,PIT-Kolbak2017}. The core issue that these techniques try to address is the label ambiguity or permutation problem (refer to Section \ref{sec:pit} for details).

In Weng et al. \cite{SingleChannelSep-Weng2015} a deep learning model was developed to recognize the mixed speech directly. To solve the label ambiguity problem, Weng et al. assigned the senone labels of the talker with higher instantaneous energy to output one and the other to output two. This, although addresses the label ambiguity problem, causes frequent speaker switch across frames. To deal with the speaker switch problem, a two-speaker joint-decoder with a speaker switching penalty was used to trace speakers. This approach has two limitations. First, energy, which is manually picked, may not be the best information to assign labels under all conditions. Second, the frame switching problem introduces burden to the decoder. 

In Hershey et al. \cite{DeepClustering-hershey2015,DeepClustering2-isik2016} the multi-talker mixed speech is first separated into multiple streams. An ASR engine is then applied to these streams independently to recognize speech. To separate the speech streams, they proposed a technique called deep clustering (DPCL). They assume that each time-frequency bin belongs to only one speaker and can be mapped into a shared embedding space. The model is optimized so that in the embedding space the time-frequency bins belong to the same speaker are closer and those of different speakers are farther away. During evaluation, a clustering algorithm is used upon embeddings to generate a partition of the time-frequency bins, i.e., speech separation and recognition are two separate components. 

Chen et al. \cite{AtrractorNet4SpeechSeparation-chen2017} proposed a similar technique called deep attractor network (DANet). Following DPCL, their approach also learns a high-dimensional embedding of the acoustic signals. Different from DPCL, however, it creates cluster centers, called attractor points, in the embedding space to pull together the time-frequency bins corresponding to the same source. The main limitation of DANet is the requirement to estimate attractor points during evaluation time and to form frequency-bin clusters based on these points.

In Yu et al. \cite{PIT-yu2017} and Kolbak et al.\cite{PIT-Kolbak2017}, a simpler yet equally effective technique named permutation invariant training (PIT) was proposed to attack the speaker independent multi-talker speech separation problem. In PIT, the source targets are treated as a set (i.e., order is irrelevant). During training, PIT first determines the output-target assignment with the minimum error at the utterance level based on the forward-pass result. It then minimizes the error given the assignment. This strategy elegantly solved the label permutation problem and speaker tracing problem in one shot. However, in these original works PIT was used to separate speech streams from mixed speech. For this reason, a frequency-bin mask was first estimated and then used to reconstruct each stream. The minimum mean square error (MMSE) between the true and reconstructed streams was used as the criterion to optimize model parameters.

In this paper, we propose the PIT-ASR model that can directly recognize multiple streams of speech given just the single-channel mixed speech, without first separating it into speech streams. Different from \cite{PIT-yu2017,PIT-Kolbak2017}, we define PIT over the cross entropy (CE) between the true and estimated senone posterior probabilities. We evaluate our approach on the artificially mixed AMI data and demonstrate that the proposed approach is very promising.

The rest of the paper is organized as follows. In Section \ref{sec:problem} we describe the speaker independent multi-talker mixed speech recognition problem. In Section \ref{sec:pit} we apply PIT-ASR to directly recognize multi-streams of speech. We report experimental results in Section \ref{sec:exp} and conclude the paper in Section \ref{sec:conclusion}.

\section{Problem Setup} \label{sec:problem}

In this paper, we assume that a linearly mixed single-microphone signal ${y}[n]=\sum_{s=1}^{S} {x}_s[n]$ is known, where ${x}_s[n], s=1,\cdots,S$ are $S$ streams of speech sources. Our goal is to separate and recognize these streams. 

However, given only the mixed speech ${y}[n]$, the problem of recognizing all streams is under-determined because there are an infinite number of possible ${x}_s[n]$ (and thus recognition results) combinations that lead to the same ${y}[n]$. Fortunately, speech is not random signal. It has patterns that we may learn from a training set of pairs $\mathbf{y}$ and ${\mathbf{\ell}_s,s=1,\cdots,S}$, where $\mathbf{\ell}_s$ is the senone label sequence for stream $s$. 

In the single speaker case, where $S=1$, the learning problem can be casted as a simple supervised optimization problem, in which the input to the model is some feature representation of $\mathbf{y}$ and the output is simply the senone posterior probability conditioned on the input. The model can be optimized to minimize the cross entropy between the senone label and the estimated posterior probability.

When $S>1$, however, it is no longer a simple supervised optimization problem due to the label ambiguity or permutation problem. Because speech sources are symmetric given the mixture (i.e., $\mathbf{x}_1+\mathbf{x}_2$ equals to $\mathbf{x}_2+\mathbf{x}_1$ and both $\mathbf{x}_1$ and $\mathbf{x}_2$ have the same characteristics), there is no pre-determined way to assign the correct target to the corresponding output layer. Interested readers can find additional information in \cite{PIT-yu2017,PIT-Kolbak2017} on how training progresses to nowhere when the conventional supervised approach is used for the multi-talker speech separation.

\section{Permutation Invariant Training} \label{sec:pit}

To address the label ambiguity problem, we propose a novel model based on the permutation invariant training (PIT) \cite{PIT-yu2017,PIT-Kolbak2017}. Note that, DPCL \cite{DeepClustering-hershey2015,DeepClustering2-isik2016} and DANet \cite{AtrractorNet4SpeechSeparation-chen2017} are alternative solutions to the label ambiguity problem when the goal is speech source separation. However, these two techniques are not suitable for direct recognition of multiple streams of speech because of the clustering step required during separation, and the assumption that each time-frequency bin belongs to only one speaker.

Formally, given some feature representation $\mathbf{Y}$ of the mixed speech $\mathbf{y}$, our model will compute

\begin{align}
  \mathbf{H}_0 &= \mathbf{Y} \\
  \mathbf{H}_i^{f} &= RNN_i^{f}(\mathbf{H}_{i-1}), i=1,\cdots,N \\
  \mathbf{H}_i^{b} &= RNN_i^{b}(\mathbf{H}_{i-1}), i=1,\cdots,N \\
  \mathbf{H}_i &= Stack(\mathbf{H}_i^{f}, \mathbf{H}_i^{b}), i=1,\cdots,N \\
  \mathbf{H}_o^s &= Linear(\mathbf{H}_N), s=1,\cdots,S \\
  \mathbf{O}^s &= Softmax(\mathbf{H}_o^s), s=1,\cdots,S
\end{align}
using a deep bidirectional recurrent neural network (RNN), where $\mathbf{H}_0$ is the input, $\mathbf{H}_i, i=1,\cdots,N$ is the $i$-th hidden layer in an $N$-hidden-layer network, $RNN_i^{f}$ and $RNN_i^{b}$ are the forward and backward RNNs at hidden layer $i$, respectively, $\mathbf{H}_o^s, s=1,\cdots,S$ is the excitation at output layer for each speech stream $s$, and $\mathbf{O}^s, s=1,\cdots,S$ is the output layer for stream $s$. Note that, in this model each output layer represents an estimate of the senone posterior probability for a speech stream. No additional clustering or speaker tracing is needed. Although various RNN structures can be used, in this study we used long short-term memory (LSTM) RNNs. It's clear that nothing is special in the forward computation.

\begin{figure}
\centering
\includegraphics[width=\linewidth]{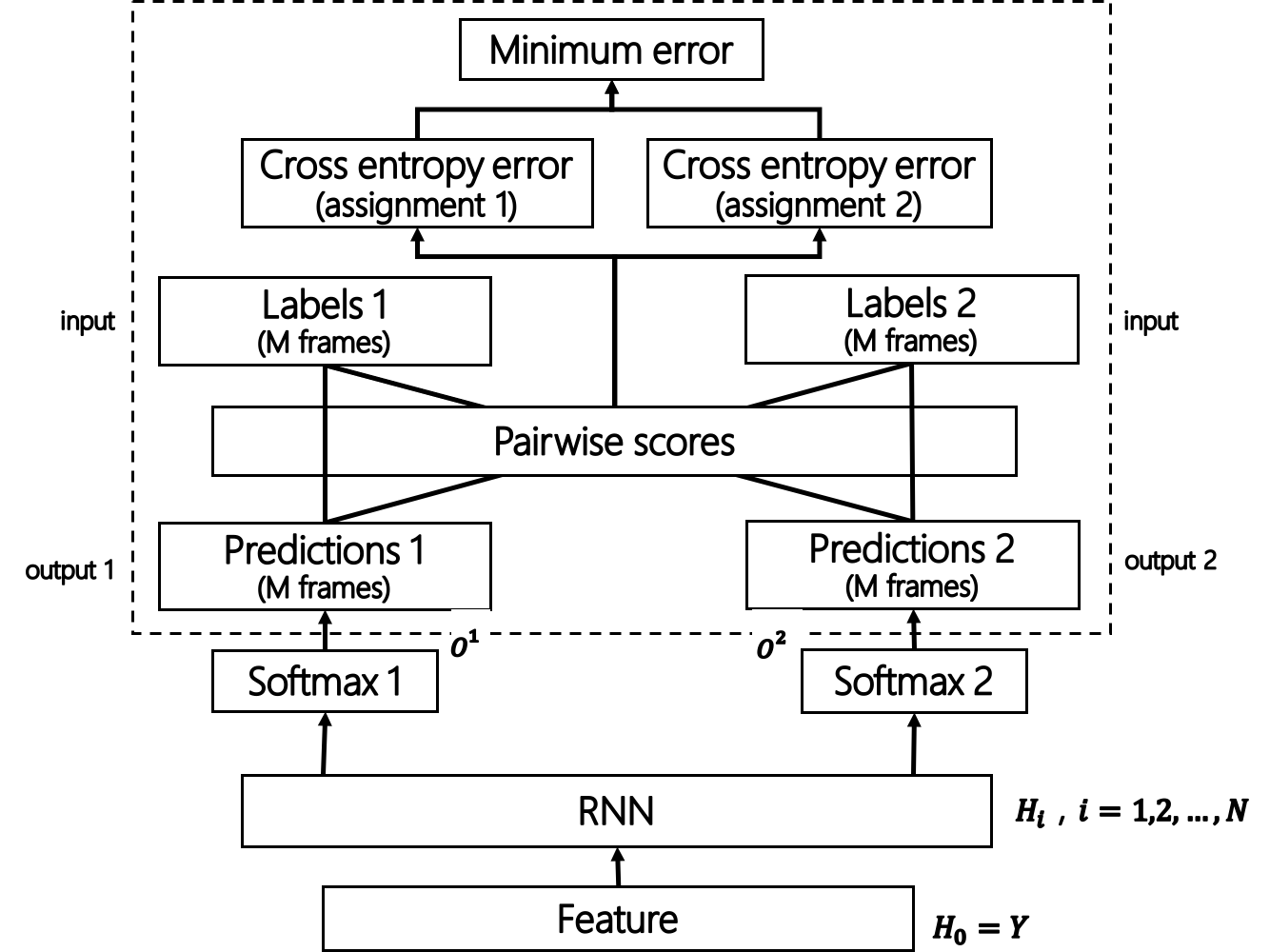}
\caption{The two-talker speech recognition model with permutation invariant training}
\label{fig:pitmodel}
\end{figure}

The key is in the training process. We need to assign the correct label to each output layer for each training sample, i.e., deal with the label ambiguity problem, and to make sure the posterior probability for the same speaker is always associated with the same output layer across frames. PIT \cite{PIT-yu2017,PIT-Kolbak2017}, which is originally designed for speech separation, is extended here to guarantee these properties. More specifically, we minimize the objective function

\begin{align}
  J &= \frac{1}{S} \min_{s' \in permu(S)} \sum_{s}  \sum_{t} { CE(\mathbf{\ell}_t^{s'},\mathbf{O}_t^s)}, s=1,\cdots,S
\end{align}
where $permu(S)$ is a permutation of ${1,\cdots, S}$. The model is illustrated in Figure \ref{fig:pitmodel}. We note two important ingredients in this objective function. First, we compute the average CE for each possible assignment of labels, pick the one with minimum CE, and optimize for that assignment. In other words, it automatically finds the appropriate assignment no matter how the labels are ordered. Second, the CE is computed over the whole sequence for each assignment. This forces all the frames of the same speaker to be aligned with the same output layer. This strategy elegantly solves the label permutation problem and speaker tracing problem in one shot. Note, the computational cost associated with label assignment is negligible compared to the network forward computation during training, and no label assignment (thus no cost) is needed during evaluation. 


\section{Experimental Results} \label{sec:exp}

To evaluate the proposed approach, a series of experiments were performed on an artificially mixed AMI corpus, and only two-talker mixed scenario is focused here.

\subsection{Experimental data}

\def \fig {figure/}
The AMI IHM (close-talk) data is used, which contains about 80 hours and 8 hours in training and evaluation sets respectively \cite{hain2012transcribing,swietojanski2013hybrid}, and the two-talker mixed speech is artificially generated with the sentences in the corpus. For the better and clear definition, we defined high energy (High E) and low energy (Low E) speakers within each two-talker mixed speech, and thus generated five different SNR conditions (i.e. 0dB, 5dB, 10dB, 15dB, 20dB) based on the energy ratio of the two-talkers. We set a rule to make the length of the selected mixed speech pair comparable, so most speech duration in this new corpus is two-talker overlapped. All the utterance-pairs are randomly chosen from two different speakers, and the shorter one will be padded with small noise at the front and end to get the same length as the longer one.

\begin{figure}[htb!]
  \centering   \includegraphics[width=0.9\linewidth]{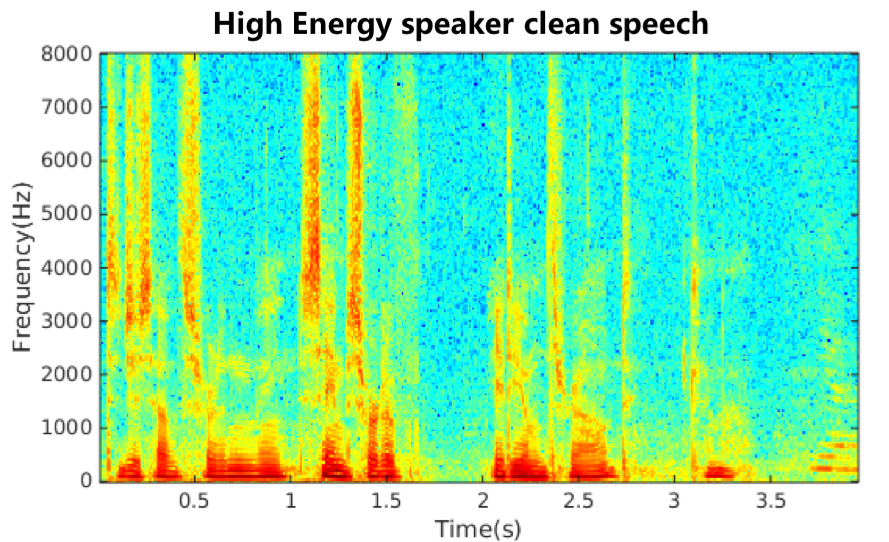}   \includegraphics[width=0.9\linewidth]{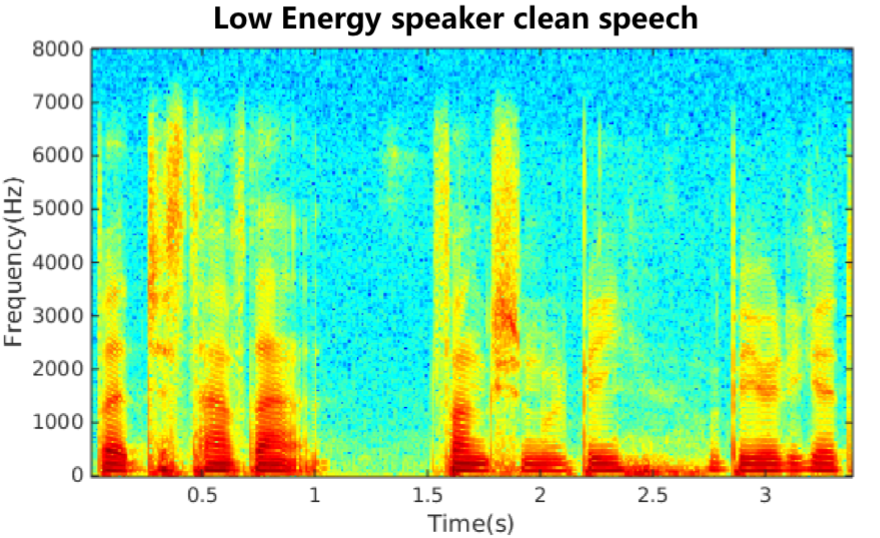}   \includegraphics[width=0.9\linewidth]{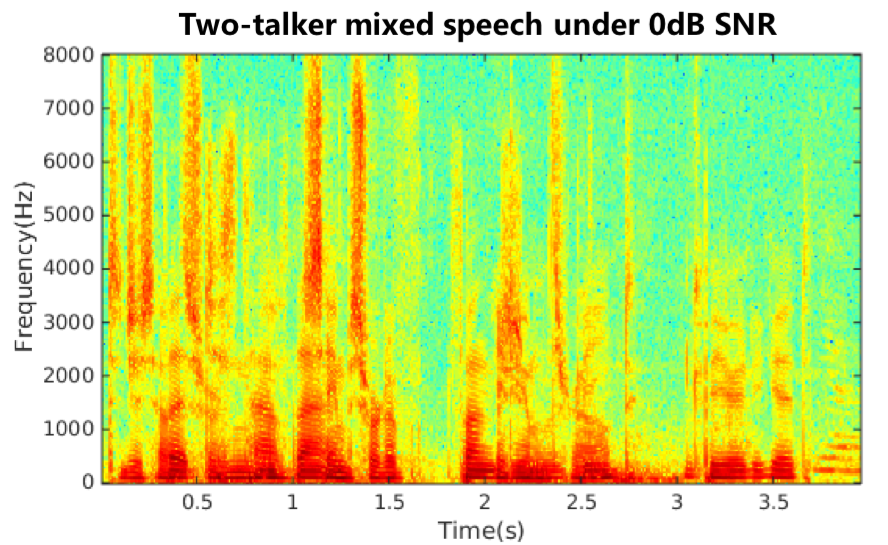}    
    \caption{Spectrogram comparison of original single-talker clean speech and the 0db two-talker mixed-speech in the new AMI corpus}
    \label{fig:spectrum}
\end{figure}

\begin{figure*}[htb!]
  \centering
  \includegraphics[width=\linewidth]{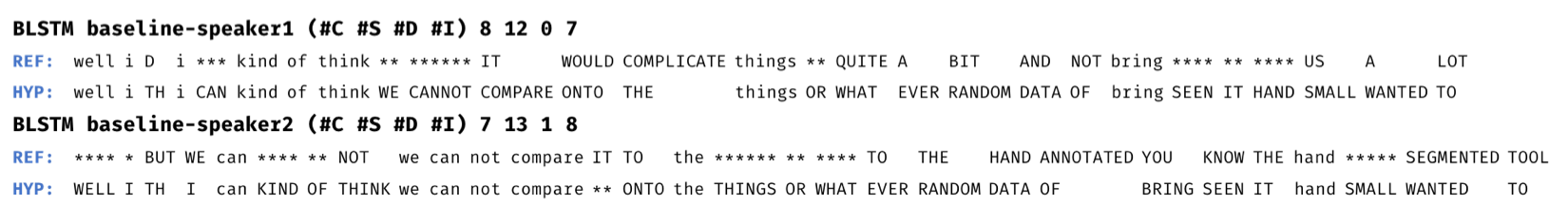}
  \caption{Decoding results of baseline BLSTM-RNN system on 0db two-talker mixed speech sample}
  \label{fig:decodeSampleBs}
\end{figure*}
\begin{figure*}[htb!]
  \includegraphics[width=\linewidth]{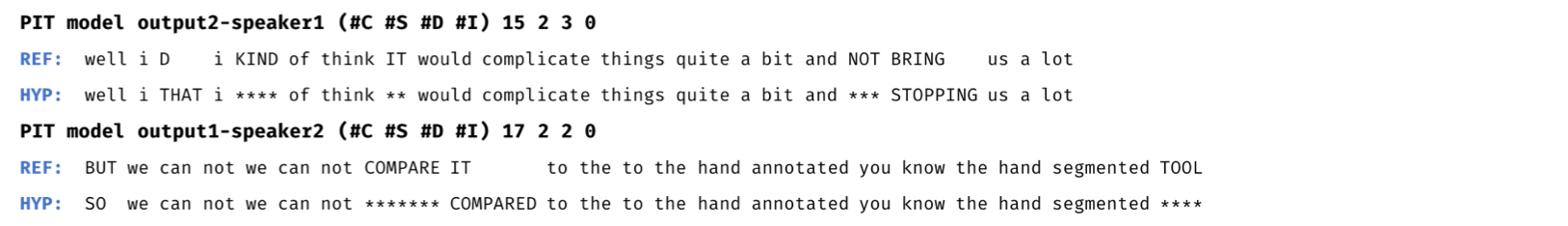}
  \caption{Decoding results of the proposed PIT-ASR model on 0db two-talker mixed speech sample}
  \label{fig:decodeSamplePIT}
\end{figure*}

Figure \ref{fig:spectrum} gives one spectrogram comparison of the original single-talker clean speech and the 0db two-talker mixed-speech in this new AMI corpus. It is observed that there is a huge difference within the two-talker and single-talker spectrograms. Two clean signals are sufficiently overlapped in the mixed speech and hard to separate them from each other. 

\subsection{Baseline setup}

In this work, all the neural networks were built using the latest Microsoft Cognitive Toolkit (CNTK) \cite{yu2014introduction} and the decoding systems were built based on Kaldi \cite{povey2011kaldi}. We first followed the officially released kaldi recipe to build an LDA-MLLT-SAT GMM-HMM model. This model uses 39-dim MFCC feature and has roughly 4K tied-states and 80K Gaussians. We then used this acoustic model to generate the senone alignment for neural network training. We trained the DNN and BLSTM-RNN baseline systems with the original AMI IHM data. 80-dimensional log filter bank features with CMVN were used to train the baselines. The DNN has 6 hidden layers each of which contains 2048 Sigmoid neurons. The input feature for DNN contains 11 frames contextual window. The BLSTM-RNN has 3 bidirectional LSTM layers which are followed by the soft-max layer. Each BLSTM layer has 512 memory cells. The input to the BLSTM-RNN is a single acoustic frame. All the models explored here are optimized with cross-entropy criterion. The DNN is optimized using SGD method with 256 minibatch size,  and the BLSTM-RNN is trained using SGD and BPTT with 4 full-length utterances parallel processing.

For decoding, we used a 50K-word dictionary and a trigram language model interpolated from the ones created using the AMI transcripts and the Fisher English corpus. The performance of these two baselines on the original single-speaker AMI corpus are presented in Table \ref{tab:baselinewerami}, and they are still comparable with other works \cite{swietojanski2013hybrid} even without using adapted fMLLR feature. It is noted that adding more BLSTM layers did not show substantial WER reduction in the baseline.

\begin{table}[th]
  \caption{WER (\%) of the baseline systems on original AMI IHM single-talker corpus}
  \label{tab:baselinewerami}
  \centering
  \begin{tabular}{ c  c  c }
    \toprule
    \textbf{Model} & \textbf{WER}  \\
    \midrule
    DNN            & 28.0             \\
    BLSTM          & 26.6               \\
    \bottomrule
  \end{tabular}
  
\end{table}

To test the baseline results on the two-talker mixed speech, the above baseline BLSTM-RNN model is utilized to decode the mixed speech directly. In scoring we compare the decoding outputs with the individual reference of two speakers respectively to obtain two-talkers' WERs, and the results are illustrated in Table \ref{tab:baselinewermixed}. It is observed that the ability of the single-speaker model is very limited on the multi-talker mixed speech, and there is very large degradation in all conditions. The performance drop is increased very fast with the lower SNR, and the WERs for the low energy speaker even are all around 100.0\%. These results demonstrate the huge challenge of the multi-talker speech recognition.

\begin{table}[th]
  \caption{WER (\%) of the baseline BLSTM-RNN system on two-talker mixed AMI IHM speech}
  \label{tab:baselinewermixed}
  \centering
  \begin{tabular}{ c  c  c }
    \toprule
    \textbf{SNR Condition} & \textbf{High E Spk} & \textbf{Low E Spk}  \\
    \midrule
    0db           & 85.0  & 100.5               \\
    5db           & 68.8  & 110.2             \\
    10db           & 51.9  & 114.9               \\
    15db           & 39.3  & 117.6               \\
    20db           & 32.1  & 118.7               \\
    \bottomrule
  \end{tabular}
  
\end{table}

\subsection{Evaluation on PIT-ASR models}

The experimental results on the proposed PIT-ASR model is described here. All the mixed data under the different SNR conditions are pooled together for training. The individual senone alignments for the two-talkers in each mixed speech utterance are from the single-speaker baseline alignment. For compatibility, the alignment of the shorter utterance within the mixed speech is padded with the silence state at the front and the end. The PIT-ASR model is composed of 4 bidirectional LSTM layers with 768 memory cells in each layer, and 40-dimensional log filter bank feature is used for the PIT-ASR model. The model was trained with 8 parallel utterances in the same minibatch, and the gradient was clipped with the threshold of 0.0003 to guarantee the training stability.


Two outputs of the PIT-ASR model are both used in decoding to obtain the hypotheses for two talkers. For scoring, we evaluated the hypotheses on the pairwise score mode against the two references, and made the better WER as the final assignment for each utterance.

The results are shown in Table \ref{tab:pitwer}. The PIT-ASR model achieves very large gains on both talkers compared with baseline results in Table \ref{tab:baselinewermixed} for all SNR conditions. The degradation increases slowly with the lower SNR for the high energy speaker, and the improvement is huge for the low energy speaker. In 0dB SNR scenario, the performances of two speakers are very close, and obtain 40.0\% relative improvement for both high and low energy speakers. In 20dB SNR, the WER of the high energy speaker is still significantly better than the baseline, and even approaches the original single-speaker decoding in Table \ref{tab:baselinewerami}.  

\begin{table}[th]
  \caption{WER (\%) of the propsoed PIT-ASR model on two-talker mixed AMI IHM speech}
  \label{tab:pitwer}
  \centering
  \begin{tabular}{c  c  c }
    \toprule
    \textbf{SNR Condition} & \textbf{High E WER} & \textbf{Low E WER}  \\
    \midrule
    0db           & 49.74  & 56.88               \\
    5db           & 40.31  & 60.31               \\
    10db           & 34.38  & 65.52               \\
    15db           & 31.24  & 73.04               \\
    20db           & 29.68  & 80.83               \\
    \bottomrule
  \end{tabular}
\end{table}

To give a better understanding on the results comparison, the results of one 0db two-talker mixed speech utterance using different models are illustrated in Figure \ref{fig:decodeSampleBs} and \ref{fig:decodeSamplePIT}. For the baseline using BLSTM-RNN decoding with the mixed speech directly, the hypotheses are erroneous and most outputs are wrong. In contrast, lots of words can be recognized correctly by the proposed PIT-ASR model for both speakers, and it seems that the PIT-ASR framework can do the speech separation implicitly. 

\section{Conclusion} \label{sec:conclusion}

In this paper, we proposed a novel technique for direct recognition of multiple speech streams given the single channel of mixed speech, without first separating them. Our technique is based on permutation invariant training, which was originally developed for separation of multiple speech streams. Our experiments on artificially mixed AMI data showed that the proposed approach is very promising. 

There are many possible ways to further improve the recognition accuracy. For example, we only explored log filter bank features. It is well known that finer frequency resolution can help better separate speech streams. In addition, we only used acoustic information in this work. Further accuracy improvement can be achieved by feeding language model information back from the decoder to the speech separation component, and by jointly considering all streams of speech when making decoding decision. Although we discussed and evaluated our proposed approach on single channel mixed speech, the technique can be applied to multi-channel condition and can exploit beam-forming results to achieve better results.

\bibliographystyle{IEEEtran}

\bibliography{mybib}

\end{document}